\documentclass[useAMS,usenatbib]{mn2e}

\usepackage{aas_macros, epsfig}
\newcommand{\gtsim}{\mbox{{\raisebox{-0.4ex}{$\stackrel{>}{{\scriptstyle\sim}}\,$}}}}

\title[Photometric redshifts using Gaussian processes]{Photometric redshift estimation using Gaussian processes}
\author[Bonfield et al.]{D. G. Bonfield$^{1}$\thanks{E-mail:
d.bonfield@herts.ac.uk (DGB)}, Y. Sun$^{2}$,
N. Davey$^{2}$,
M. J. Jarvis$^{1}$,
F. B. Abdalla$^{3}$,
M. Banerji$^{4,3}$,
\and
R. G. Adams$^{2}$
\\~\\
$^{1}$Centre for Astrophysics Research, Science \& Technology Research Institute, 
University of Hertfordshire, Hatfield, AL10 9AB, UK \\
$^{2}$Centre for Computer Science \& Informatics Research, Science \& Technology Research Institute, 
University of Hertfordshire, \\Hatfield, AL10 9AB, UK \\
$^{3}$Department of Physics \& Astronomy, University College London, London, WC1E 6BT, UK\\
$^{4}$Institute of Astronomy, University of Cambridge, Cambridge, CB3 0HA, UK}

\begin{document}

\date{}

\pagerange{\pageref{firstpage}--\pageref{lastpage}} \pubyear{2009}

\maketitle

\label{firstpage}

\begin{abstract}
We present a comparison between Gaussian processes (GPs) and artificial neural networks (ANNs) as methods for determining photometric redshifts for galaxies, given training set data.  In particular, we compare their degradation in performance as the training set size is degraded in ways which might be caused by the observational limitations of spectroscopy.  
Using publicly-available regression codes, we find that performance with large, complete training sets is very similar, although the ANN achieves slightly smaller root mean square errors.  
Training sets with brighter magnitude limits than the test data do not strongly affect the performance of either algorithm, until the limits are so severe that they remove almost all of the high-redshift training objects.  
Similarly, the introduction of a plausible number (up to 10\%) of inaccurate redshifts into the training set has little effect on either method.  
However, if the size of the training set is reduced by random sampling, the RMS errors of both methods increase, but they do so to a lesser extent and in a much smoother manner for the case of GP regression; for the example presented {\sc{ANNz}} has RMS errors $\sim20\,$\% worse than GP regression in the small training-set limit.  
Also, when training objects are removed at redshifts $1.3<z<1.7$, to simulate the effects of the ``redshift desert'' of optical spectroscopy, the Gaussian process regression is successful at interpolating across the redshift gap, while the ANN suffers from strong bias for test objects in this redshift range.    
Overall, GP regression has attractive properties for photometric redshift estimation, particularly for deep, high-redshift surveys where it is difficult to obtain a large, complete training set.  
At present, unlike the ANN code, public GP regression codes do not take account of inhomogeneous measurement errors on the photometric data, and thus cannot estimate reliable uncertainties on the predicted redshifts.  However, a better treatment of errors is in principle possible, and the promising results in this paper suggest that such improved GP algorithms should be pursued.

\end{abstract}

\begin{keywords}
galaxies: distances and redshifts -- surveys -- methods:data analysis
\end{keywords}

\section{Introduction}
\subsection{Photometric redshift estimation}
The idea of estimating galaxy redshifts from broad-band photometry was first applied by \citet{Baum62} 
 to obtain redshifts for suspected members of a cluster which were too faint for spectroscopy using the instruments of the time.  
Baum determined the redshifts by first measuring photometry of a set ellipticals in known redshift clusters, to create a coarsely-sampled template spectral energy distribution, and then shifting this in logarithmic wavelength space (as if redshifted) until it matched the photometry of the unknown objects.

Since then it has become possible to fit to a wide variety of higher resolution templates, based on empirical \citep[e.g.][]{cww} or synthetic \citep[e.g.][]{bc03} galaxy spectra.  Model photometry is obtained by directly multiplying a redshifted template spectrum by the measured response curves for each photometric band, and is compared with data to determine the best-fitting redshift.  Template-fitting methods are very widely used \citep[e.g.][]{hyperz,bpz}, since they can be applied to any photometric dataset provided that the instrument response curves are known and the templates appropriate for the galaxy being studied.  However, there are often systematic uncertainties in both of these; filter curves, detector responses and atmospheric transmission are in general not precisely known, and spectral templates are necessarily composed of or calibrated against low-redshift galaxies, and become less reliable as redshift increases.  

Some of these problems with template fitting can be mitigated by using known redshifts for calibration \citep[e.g.][]{ilbert06,feldmann2006}.
However, where a representative subset of the objects have precisely known redshifts, it is much simpler to use these objects to directly train some function so that it returns redshift as a function of photometric data, and then apply this function to the remainder of the catalogue.  Empirical training-set methods of this kind have the additional advantage that it is straightforward to include information from additional catalogued parameters such as angular size or surface brightness profile.  

Empirical photometric redshift methods are currently dominated by artificial neural networks \citep[ANNs;][]{firth2003,Vanzella2004}.  
ANNs can perform general non-linear mappings (e.g. from photometric data to redshift) by the use of multiple layers of linear combinations of data (optionally with non-linear transfer functions on some of the nodes), with the weights on these combinations being free parameters which are modified during training.  Simple feed-forward networks (in the sense that values propagate in one direction only, e.g. from photometry, via intermediate layers of linear combinations, to the estimated redshift) have the useful property, during training, that the errors on the redshift estimates help inform the direction in which to modify the weights, and they can thus be trained quite efficiently.  

The most commonly-used implementation of ANNs for redshift estimation is the publicly-available {\sc{ANNz}} package \citep{annz04}.  \citet{megazlrg} show that {\sc{ANNz}}  matches or surpasses the performance of the best publicly-available template-fitting codes, when it is provided with a high quality training set.  However, neural networks are far from the only tools able to perform non-linear regression. One alternative, which has recently been favoured in the machine-learning community, is a class of models known as Gaussian processes.

\subsection{Gaussian process regression}
A Gaussian process (GP) is defined simply as a collection of random variables which have a joint Gaussian distribution.  The utility of GPs for regression is that they can be used as a prior distribution over a space of functions which are considered as possible models for the data \citep[e.g.][]{wr96}.  The space of functions represented by a Gaussian process is completely defined by a mean function (usually taken to be zero) and a covariance function, which for $n$ training data ${\bf{x}}_1...{\bf{x}}_n$ (each of which would be a vector of photometry in our case) and one test point ${\bf{x}}_*$ (i.e. the photometry of a galaxy with unknown redshift) takes the form of an $(n+1) \times (n+1)$ matrix of covariances $k_{ij}=k({\bf{x}}_i,{\bf{x}}_j)$ between the data.  

Thus, the vector of ``outputs'', which contains redshifts for the training data $y_1...y_n$ and the redshift to be inferred for the test point $y_*$, is assumed to be given by:

\begin{equation} \label{eq:smalldef}
\left( \begin{array}{c}
y_1 \\ \vdots \\ y_n \\ y_*
\end{array} \right) = 
\mathcal{N}\left({\bf{0}}, \left[ \begin{array}{cccc}
k_{11} + \sigma_n^2& \cdots & k_{1n} & k_{1*} \\
\vdots & \ddots & & \vdots \\
k_{n1} &  &  k_{nn} +\sigma_n^2 & k_{n*} \\
k_{*1} & \cdots &  k_{*n} & k_{**} 
\end{array} \right] \right)
\end{equation}

where $\mathcal{N}({\bf{0}},C)$ is a joint Gaussian distribution with mean ${\bf{0}}$ and covariance matrix $C$, and $\sigma_n^2$ is a noise term added to the first $n$ diagonal elements, both to account for noise in the data (assumed independent and identically distributed) and to increase numerical stability.  

While this distribution appears extremely simple, it should be noted that there is considerable freedom to choose the covariance function $k({\bf{x}}_i,{\bf{x}}_j)$, also commonly referred to as the kernel, which can be any positive definite function.  This means we can achieve non-linear regression by using a non-linear kernel.
We will discuss our specific choice of kernel later, but in general one chooses some function whose value indicates the similarity of datapoints ${\bf{x}}_i$ and ${\bf{x}}_j$, and which may have a number of hyperparameters whose optimisation forms part of the training process.  

If we define
\begin{displaymath}
K = 
\left( \begin{array}{ccc}
k_{11} & \cdots & k_{1n} \\
\vdots & \ddots &  \\
k_{n1} &  &  k_{nn} 
\end{array} \right)
,
\,{\bf{k}}_* = 
\left( \begin{array}{c}
k_{1*} \\
\vdots \\
k_{n*} 
\end{array} \right),
{\bf{y}} = 
\left( \begin{array}{c}
y_{1} \\
\vdots \\
y_{n} 
\end{array} \right)
\end{displaymath}
then it can be shown
\citep[e.g.][]{rw06} that the posterior probability distribution of the predicted redshift $y_*$ is a Gaussian distribution, with mean $\hat{y}_*$ and variance $\sigma_*^2$ given by
\begin{eqnarray}\label{eq:gprmean}
\hat{y}_* & = & {\bf{k}}_*^T(K+\sigma_n^2I)^{-1}{\bf{y}} \\
\label{eq:gprvar}
\sigma_*^2 & = & k_{**} - {\bf{k}}_*^T(K+\sigma_n^2I)^{-1}{\bf{k}}_*
\end{eqnarray}
where $A^T$ denotes the transpose of $A$, $A^{-1}$ denotes the inverse of $A$, and $I$ is the identity matrix.

\subsection{Previous work}

\citet{Way06} first evaluated Gaussian processes for photometric redshift estimation, and found that an ensemble of  
neural networks produced a slightly smaller RMS error.  However, they used a smaller 
training set for the GP-based method due to computational limitations (the na\"{\i}ve implementation of equations \ref{eq:gprmean} and \ref{eq:gprvar} requires the inversion of an $n\times n$ matrix, which involves $O(n^3)$ operations) so this was not a completely fair comparison of the methods.  

More recently, \citet{foster09} have shown that it is possible to use rank-reduction techniques which allow inference using large training sets, with $n$ objects, to be performed in $O(nm^2)$ operations (where $m<n$).  
Such methods make use of a full covariance matrix for a subset of only $m$ objects (an $m\times m$ matrix), in conjunction with an $m\times n$ matrix of covariances between the full and partial training sets.  

With the ability to use information from a large training set, of the same size as used by ANN-based methods, \citet{Way09} find that GP regression yields a slightly lower RMS error than {\sc{ANNz}}, when applied to galaxies drawn from the Sloan Digital Sky Survey \citep[SDSS;][]{sdss2000}.  Way et al.~also characterise the performance of GP regression using different kernel functions and choices for the size of the reduced rank $m$.  

These existing evaluations of GP regression for photometric redshift estimation have used an SDSS-type survey as a model.  That is, they have assumed that a very large set of training objects is available, and have implicitly ensured that the training set is absolutely complete by selecting training and test objects in the same way from the SDSS spectroscopic sample (i.e. those SDSS galaxies which have known spectroscopic redshifts).  Use of SDSS data has also resulted in the exploration of a rather narrow range in redshift.

In this paper, we extend this work by considering a number of observationally-motivated restrictions on the training sets.  In particular, we consider cases where the number of available spectra are small, but complete, and cases in which the training set is limited in magnitude in a different manner than the test data.  We also examine the effect of reducing the number training objects with spectral types lacking emission lines, or in the emission line ``desert'' at redshifts $1.3<z<1.7$, and compare the resistance of the methods to ``bad'' training objects with incorrectly determined redshifts.  We use photometry simulated to represent a deep, high-redshift survey.

\section{Method}

\subsection{Algorithms}
For all tests of GP regression in this paper, we use the stable, reduced-rank GP regression code developed by \citet{foster09} and made publicly available as the {\sc{stableGP}} package\footnote{https://dashlink.arc.nasa.gov/algorithm/stablegp/} for {\sc{MATLAB}}\footnote{http://www.mathworks.com/}.  For simplicity, we consider only one of the possible GP algorithms available in this package, 
the ``SR-VP'' method, since it is considered to be the most numerically stable and accurate technique.  We set the reduced rank $m=800$, since the results of Way et al.~show that the precision of the method is a weak function of $m$, and they find good performance with this rank size.  
The GP kernel used is the ``neural network'' covariance function.  This function is so named because it was shown by \citet{neal96} to be equivalent to a feed-forward neural network with a single hidden layer, in the limit of an infinite number of hidden units.  It can be written
\begin{equation}\label{eq:knn}
k_{nn}({\bf{x}}_p,{\bf{x}}_q) 
= 
\alpha \, \mathrm{sin}^{-1}\left( \frac{\tilde{{\bf{x}}}_p^T \beta I \tilde{{\bf{x}}}_q}{\sqrt{(1+  \tilde{{\bf{x}}}_p^T \beta I \tilde{{\bf{x}}}_p )(1+\tilde{{\bf{x}}}_q^T \beta I \tilde{{\bf{x}}}_q)}} \right)
\end{equation}
where $\alpha$ and $\beta$ are scalar hyperparameters to be optimised, and $\tilde{{\bf{x}}}$ is the ${\bf{x}}$ vector extended by appending an element with the value 1.

We compare the performance of the GP regression method against that of the {\sc{ANNz}} neural network code.  In {\sc{ANNz}} we use a network architecture with two hidden layers, each with twice the number of nodes as we have inputs, which was shown to be effective for photometric data by \citet{annz04} and has become the default option for most users of {\sc{ANNz}}.  Since our simulated dataset has 7 photometric bands, this means a $7:14:14:1$ network. We use precisely the same training data with {\sc{ANNz}} and the GP method. However, {\sc{ANNz}} uses an additional set of objects with known redshifts as a ``validation'' dataset, which
helps prevent the network from becoming over-specialised on the objects in the training set.  Over-specialisation, which means that the model fits the training data in such detail that it performs poorly on unseen data, is a particular problem with small training sets.  We note that it is possible, in principle, to train ANNs  without the use of a validation set by using a Bayesian framework  \citep[e.g.][]{mackaythesis}.

Our GP implementation is extremely simple, and in its current form does not make use of the estimated errors on the photometry of the training or test data.  As such, it does not provide realistic estimates of the errors in derived redshifts, which are provided by {\sc{ANNz}} and can be used to define a ``clean'' sample of objects with reliable redshift estimates as shown by \citet{abdalla2008}.  In the presentation of our ANN results, we show both the full set of test objects and a set ``cleaned'' by the removal of objects with predicted errors of $\sigma_z>0.3$, so that the effectiveness of this procedure can be judged.  
In future work, we hope to provide reliable estimates of errors in quantities predicted by GP regression, and are exploring options such as monte-carlo realisations of training and/or test catalogues in order to more fully account for measurement errors.  

\subsection{Simulated data}
We use a photometric dataset which was designed to simulate the performance of a proposed configuration for the DUNE dark energy mission (since redesigned and incorporated into the Euclid mission\footnote{http://sci.esa.int/science-e/www/area/index.cfm?fareaid=102}), with infrared $J$ and $H$ filters plus a very broad $RIZ$ filter, in combination with a ground-based optical survey with the $g$, $r$, $i$, and $z$ filters, to depths proposed in an early DES (Dark Energy Survey\footnote{https://www.darkenergysurvey.org/}) configuration.  This dataset is part of a larger set of simulations used by \citet{abdalla2008} to evaluate the effects of filter selection and training set completeness on the accuracy of photometric redshifts calculated using {\sc{ANNz}}.  

The catalogue that we use consists of 142803 objects, flux limited at $RIZ < 25$.  We first split this into training, validation (used only with {\sc{ANNz}}), and test datasets, with 23638, 11949, and 107216 objects respectively.  We refer to these training (validation) data as the ``full training (validation) set''.  Throughout this work we use the complete set of test data, to represent the galaxies which would be observed photometrically in a potential future survey such as this one.  
To examine the possible effects of incomplete spectroscopic information (something which is difficult to avoid in imaging surveys which probe faint objects at high redshift), we restrict the training and validation sets in several ways.  

Our first test is to restrict the training set to brighter flux limits than the test data.  This is motivated by the fact that, for a given telescope aperture, it is generally possible to image fainter objects than one can obtain spectra for in a reasonable amount of observing time.  Placing magnitude limits of $RIZ < 23$ and $RIZ < 22$ on the training and validation sets results in  training sets with 2721 and  858 objects, and  validation sets with 1410 and 414 objects respectively.  
%

We also test the effects of training set incompleteness due to missing objects of a particular type or redshift range, since spectroscopic redshift surveys often have better completeness for objects with easily detectable emission-lines.  To do this we test a series of training sets with different fractions (20\%, 40\%, 60\%, 80\%, and 100\%) of the early-type galaxies (i.e. those simulated using E or E+S0 templates) removed randomly, and similarly test the removal of galaxies (of all spectral types) in the $1.3<z<1.7$ ``redshift desert'', where no strong emission lines are accessible to optical spectrographs.  

Our third test is to randomise the redshifts of a fraction of training set objects.  We do this because spectroscopic samples quite often contain a small fraction of objects with incorrectly determined redshifts \citep[e.g.][]{FS2001}.  

The final test uses a ``complete'' training set, with the same magnitude limit and selection effects as the full training set, but reduces the number of objects by drawing objects at random from this set.  We use randomly selected training sets containing 100, 200, 500, 1000, 2000, 5000, and 10000 objects.  For {\sc{ANNz}} we also produce random validation sets with half as many objects as the training sets.

\section{Results and discussion}

\begin{figure*}
\includegraphics[angle=0,width=0.88\textwidth,clip]{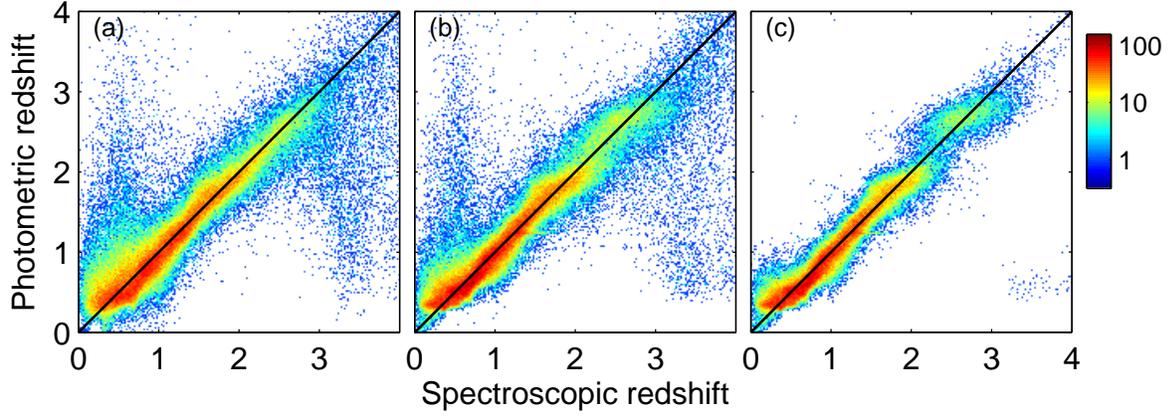}
\caption{Photometric redshifts recovered using (a) GP regression and (b) ANNz, trained with the full set of 23638 training objects (and an additional 11949 objects for validation with ANNz).  Panel (c) shows the ANNz results ``cleaned'' by the removal of test objects with estimated redshift errors $> 0.3$.  The colour scale indicates the number of test objects in a pixel, where pixels represent intervals of 0.02 in redshift.}\label{fig:zzfull}
\end{figure*}

\begin{figure*}
\includegraphics[angle=0,width=0.88\textwidth,clip]{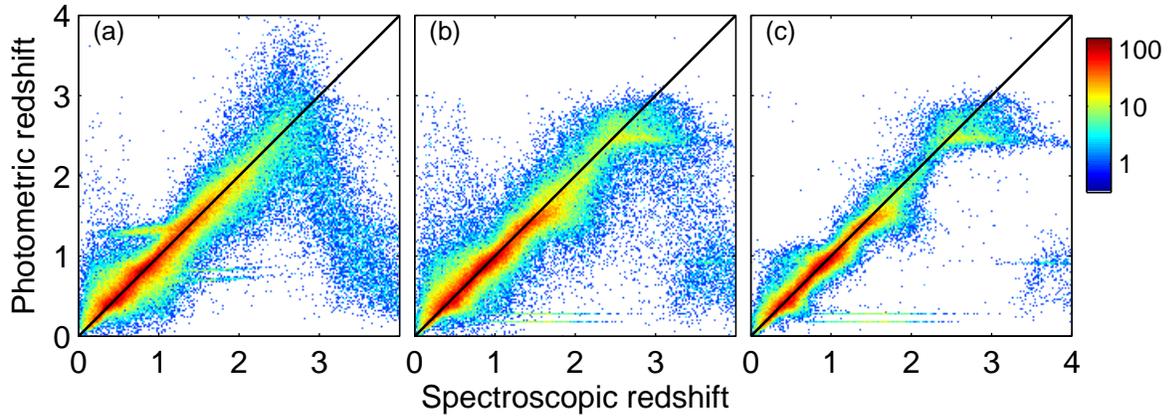}
\caption{As Figure \ref{fig:zzfull} but using only the 2721 training objects (and 1410 validation objects) with $RIZ<23$.}\label{fig:zz23}
\end{figure*}

\begin{figure*}
\includegraphics[angle=0,width=0.88\textwidth,clip]{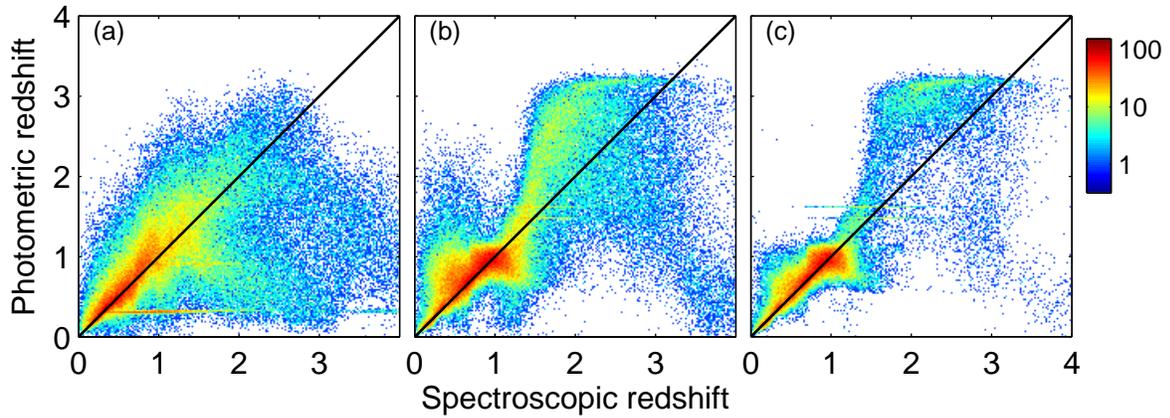}
\caption{As Figure \ref{fig:zzfull} but using only the 858 training objects (and 414 validation objects) with $RIZ<22$.}\label{fig:zz22}
\end{figure*}

Figure \ref{fig:zzfull} compares the photometric redshift estimates from GP regression and {\sc{ANNz}} when trained using the full $RIZ<25$ training set.   We show the estimated photometric redshift as a function of spectroscopic (true) redshift, using colour to indicate the density of points.  For {\sc{ANNz}} we also show the distribution obtained after removing all points with estimated uncertainties $\sigma_z>0.3$ (in this case 30\% of the points are removed); for some applications it is preferred to have an incomplete sample of galaxies such as this, with more reliable and precise photometric redshifts, but for many applications removal of such objects limits the scientific usefulness of the sample. For example the study of galaxy luminosity functions, evolution of galaxies with redshift and measurements of environmental density would be severely hampered by this kind of incompleteness. 

When using the full training set, the ANN-based method has a slightly tighter correlation, particularly at low redshift, but both methods perform quite well, with similar outliers.  While this comparison is not especially discriminatory, it demonstrates that GPs, like ANNs, are able to cope well with a larger range in redshift than present in current large surveys such as the SDSS.  Computationally, the methods are comparable even with the full training set, each taking a few minutes to train on a fairly standard workstation; this would not be the case if a na\"{\i}ve, full GP regression was used instead of the reduced-rank technique.  
With the reduced rank size $m$ held constant, the calculation time for GP regression increases only linearly with training set size $n$.

When the ANN results are ``cleaned'' by removing objects with estimated redshift errors $\sigma_z>0.3$ (panel (c)), many outliers are removed, improving the RMS error considerably (from 0.319 to 0.130 for objects in the range $0.5<z<1.0$ and from 0.224 to 0.159 for redshifts $1.5<z<2.0$; the RMS errors for the GP regression are 0.353 and 0.202 in the same redshift intervals).  

Figures \ref{fig:zz23} and \ref{fig:zz22} show similar comparisons  for the tests of magnitude-limited training sets, with $RIZ < 23$ and $RIZ < 22$ respectively.  
Perhaps surprisingly, the $RIZ < 23$ training set yields slightly better performance than the full set at the lowest redshifts, especially for GP regression, reducing the number of outliers and increasing the density of points close to the $y=x$ line.  This is likely due to 
the removal of objects at $z\gtsim 3$, which can have similar colours to low redshift objects due to the redshifted Lyman break at $\sim912\,$\AA{} mimicking the $\sim4000\,$\AA{} Balmer break, and thus cause ambiguity.  In general, even with this rather conservative magnitude limit we find that the performance of GPs and ANNs is similar, and, except for high-redshift objects, not a great deal worse than that with the full training set.  

It is only when we make the  $RIZ<22$ cut in the training set, three magnitudes brighter than the limit of the test catalogue, that we find a very substiantial loss in performance.  Although both methods perform poorly, it is worth noting some differences between the two techniques.  The GP regression suffers from very considerable scatter, but has  redshift estimates which are less biased on average than those from {\sc{ANNz}} (i.e. the mean estimate lies closer to the true value); this is potentially a serious issue for those studying large-scale structure, as the systematic clumping of estimated redshifts can lead to false detections of overdensities \citep[for a full discussion see, e.g.,][]{cvb07}.  It is also important to note that the ``cleaned'' sample from {\sc{ANNz}} has roughly the same uneven distribution as the full set; this makes the point that reliability of the estimated redshift errors can also be affected by incompleteness in the training set.  

Although the magnitude-limited training sets clearly alter the performance of the redshift estimation, the fact that a cut from $RIZ<25$ to $RIZ<23$ has a minimal effect suggests that the distribution of magnitudes of objects in the training set is not the crucial parameter.  The absolute number of training objects at the redshifts of interest appears to be more significant, i.e. the $RIZ<22$ training set is very poor at high redshift because it contains, for example, only 11 training objects with $z>1$.  The following results, showing the effect of reducing the training-set size by drawing a random sample of objects, demonstrate this effect more clearly.  

\begin{figure}
\includegraphics[angle=0,width=0.48\textwidth,clip]{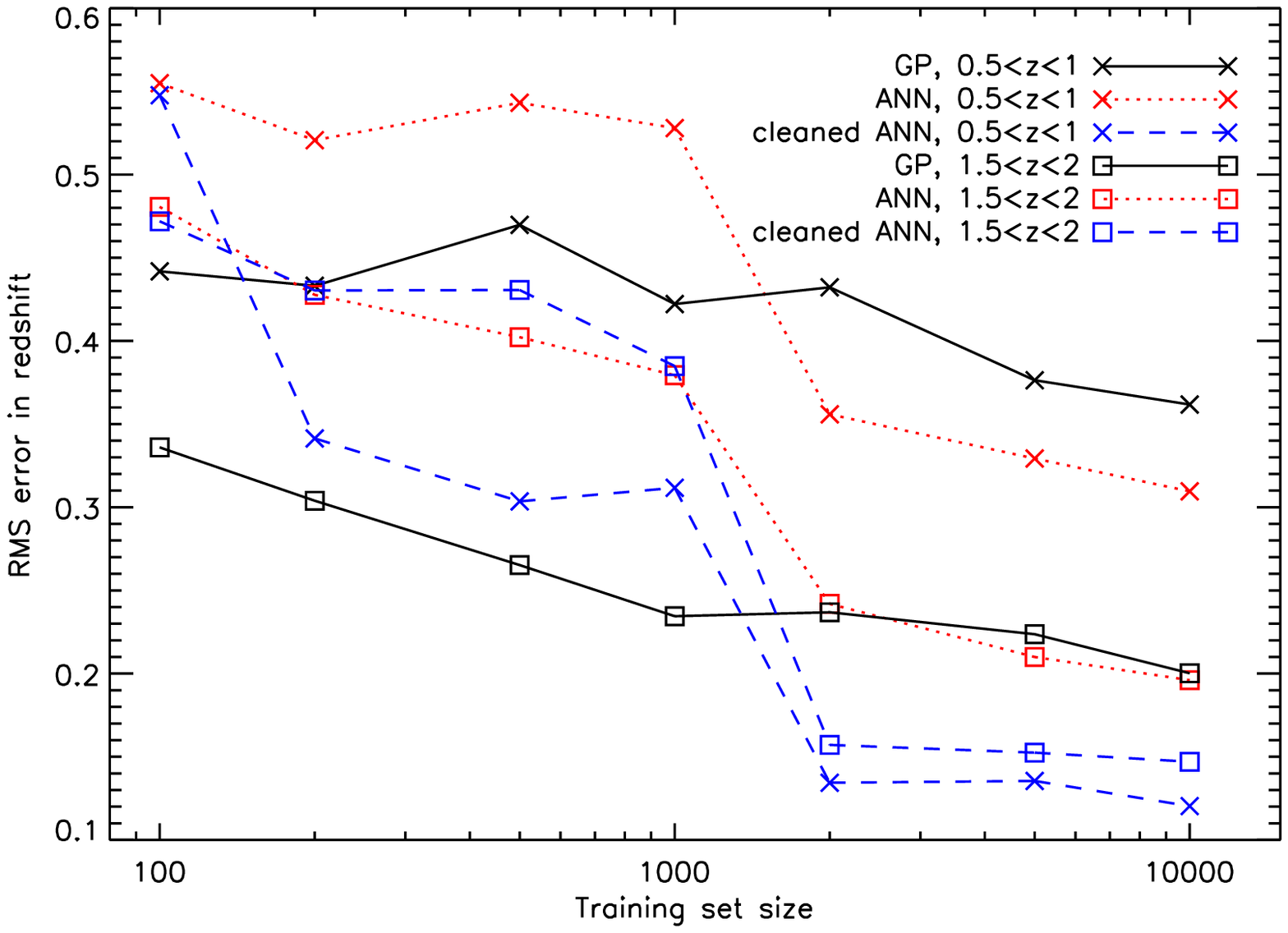}
\caption{A comparison of the RMS error in photometric redshift estimates (defined for the objects in each range in spectroscopic redshift as $\langle(z_{\mathrm{phot}}-z_{\mathrm{spec}})^2\rangle^{1/2}$) as a function of training set size, when training and validation objects are randomly selected from a large, complete set of objects with known redshifts.  
}\label{fig:sizetest}
\end{figure}

\begin{figure*}
\includegraphics[angle=0,width=0.88\textwidth,clip]{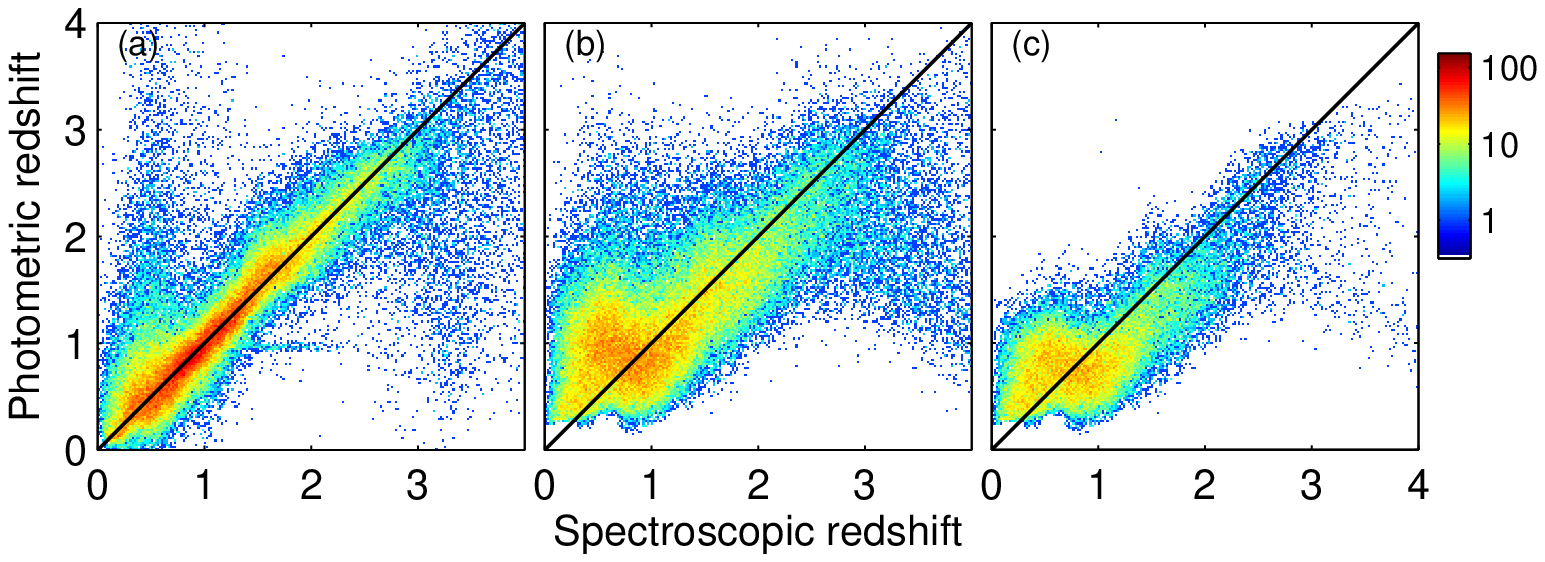}
\caption{As Figure \ref{fig:zzfull} but using 500 randomly selected training objects (and, for ANNz only, 250 randomly selected validation objects).  }\label{fig:zzrand}
\end{figure*}

Figure \ref{fig:sizetest} shows how the RMS error in photometric redshift estimates (calculated over two redshift ranges -- $0.5<z<1$ and $1.5<z<2$) changes as a function of training set size, when the training sets all have the same redshift distribution as the full set.  As discussed above, the performance in the limit of large training sets is very similar, with ANNs slightly better in the low-redshift bin and GPs slightly better at high-redshift.  However, there is a clear difference between the performance of the GP and ANN methods for small training sets.
While the RMS error increases for both methods with decreasing training-set size, there is a rather sharp transition for {\sc{ANNz}} and a much more gradual decline in the case of GP regression.  While there will clearly be difficulties, in practice, in assessing the accuracy of any photometric redshift method when only a small number of known redshifts is available, this smooth behaviour as a function of training-set size makes it appear more feasible to estimate GP regression accuracy (by, say, using half the known data for training and half for testing) than in the case of ANNs.

We choose to quote RMS errors in particular redshift ranges because the overall RMS error is not especially informative; it is too strongly effected by performance at low redshift where the majority of simulated objects lie.  Even so these RMS measures, while useful metrics, do not provide information about bias and other kinds of false structure in the redshift distributions, so for particular cases of interest we also show the detailed plots of photometric redshift versus true redshift, as for the magnitude-limited training sets.  Figure \ref{fig:zzrand} shows the detailed performance for the case with 500 training set objects (and, for the ANN method only, 250 validation objects).  It is clear from this plot that the results from {\sc{ANNz}} are more biased than those from GP regression, as well as having larger RMS error.  

For comparison, we also attempted to calculate photometric redshifts using the {\sc{HyperZ}} template-fitting code \citep{hyperz}, which does not require any training data at all.  We tried empirical and theoretical \citep{bc03} templates.  While the theoretical templates were better than the empirical ones, they still performed less well than the GP regression even with only 100 training objects, achieving RMS errors in the ranges $0.5<z<1.0$ and $1.5<z<2.0$ of 0.794 and 0.390 respectively.

We should note that one could choose a simpler neural network architecture which, with fewer free parameters, would be easier to train with a smaller training set; in general, one can optimise the network architecture to suit the size of the training set.  This could be achieved either by the use of extra validation data (which would, by definition, be in short supply in cases with few available redshifts), or model selection by optimising the Bayesian evidence \citep[e.g.][]{mackay94}.  However, while possible in principle, and indeed worth pursuing in practice, neural network architecture optimisation is not at present widely used for the calculation of photometric redshifts.  


\begin{figure}
\includegraphics[angle=0,width=0.48\textwidth,clip]{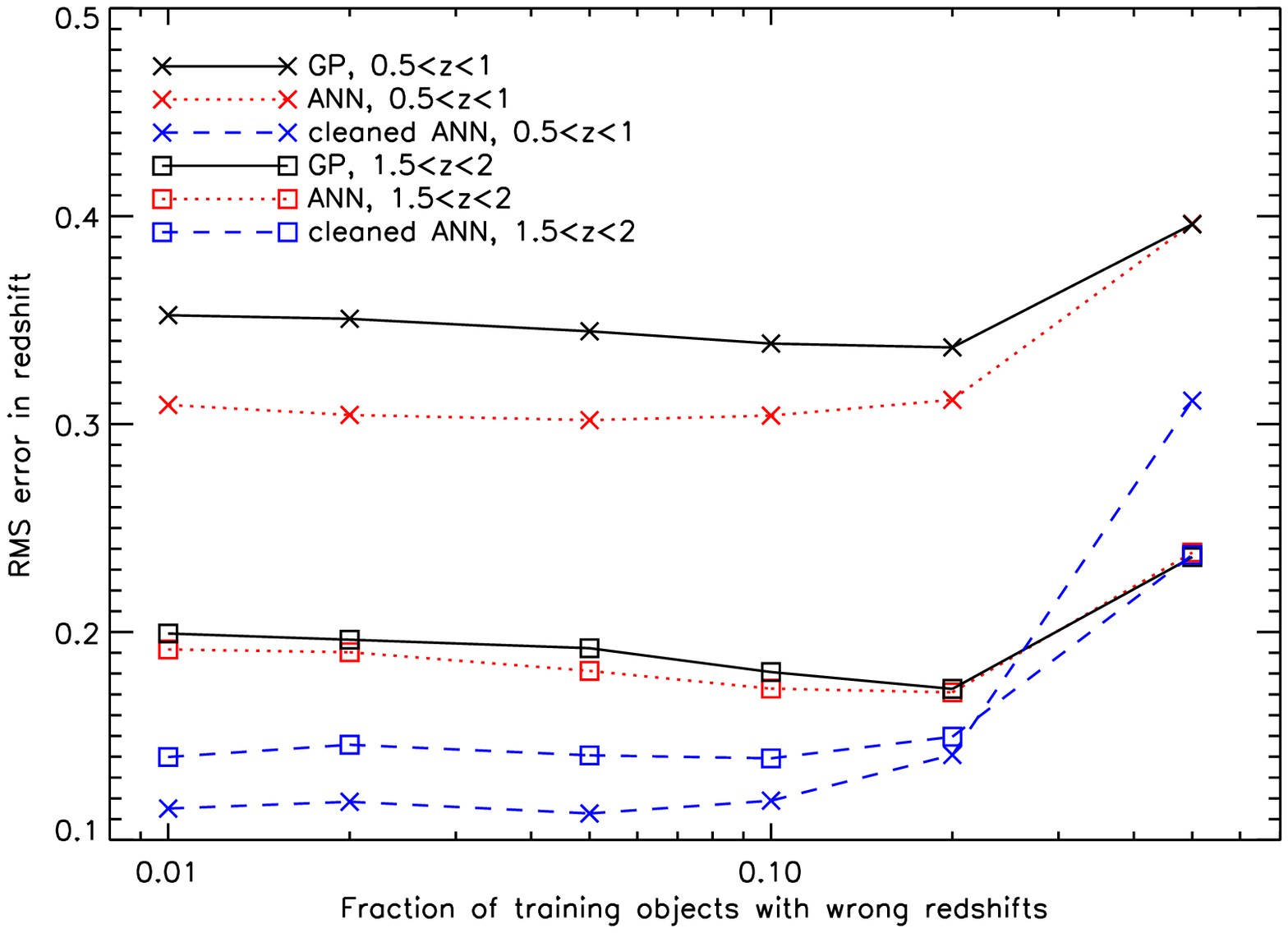}
\caption{As Figure \ref{fig:sizetest}, but showing the effect on RMS errors of randomly changing the redshifts of a subset of training objects.
}\label{fig:wrongtest}
\end{figure}

\begin{figure}
\includegraphics[angle=0,width=0.48\textwidth,clip]{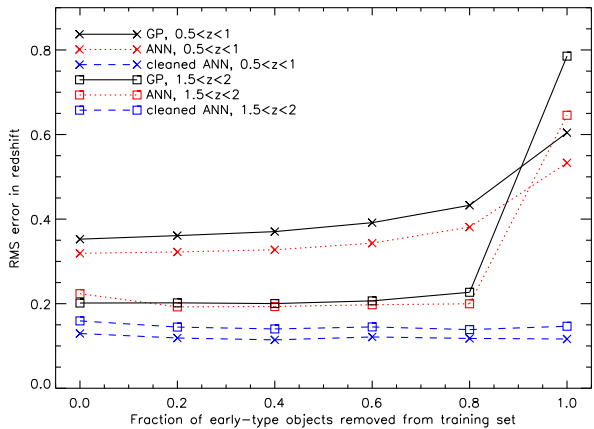}
\caption{As Figure \ref{fig:sizetest}, but showing the effect on RMS errors of removing objects with early-type template spectra.
}\label{fig:typetest}
\end{figure}

Most of the other training-set limitations tested have small effects on the RMS errors, which are similar for both ANNs and GPs.  Figure \ref{fig:wrongtest} shows that changing the redshifts of a subset of training objects (to a random, incorrect value) has little effect until the fraction of objects modified is large.  \citet{FS2001} find a wrong redshift fraction of only $\sim6\,$\% in their spectroscopic data, which appears to be too small to have an effect.   Figure \ref{fig:typetest} shows that removing objects of a particular spectral type (in this case those with some fraction of an elliptical template) also has a negligible effect until all objects of that type are removed (a situation which is unlikely to occur in practice).

\begin{figure}
\includegraphics[angle=0,width=0.48\textwidth,clip]{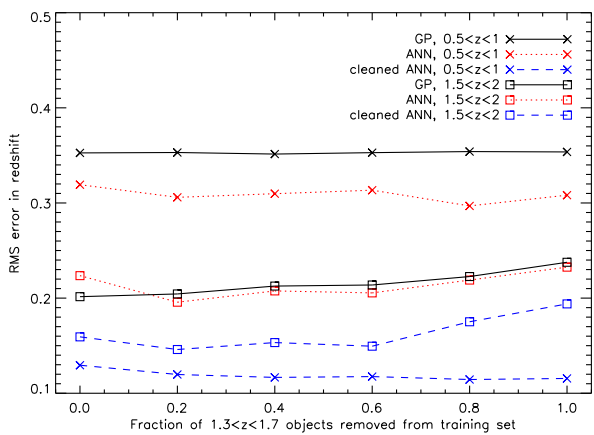}
\caption{As Figure \ref{fig:sizetest}, but showing the effect on RMS errors of removing objects with redshifts $1.3<z<1.7$.
}\label{fig:deserttest}
\end{figure}

\begin{figure*}
\includegraphics[angle=0,width=0.88\textwidth,clip]{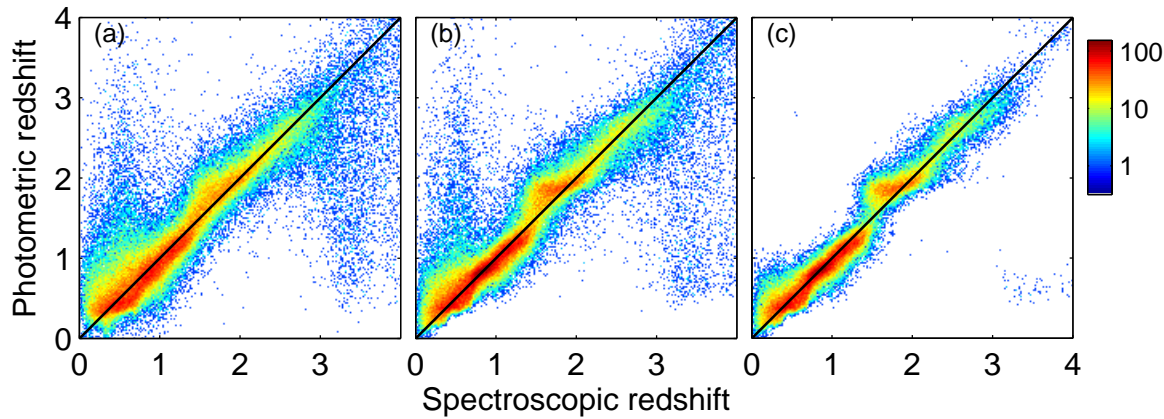}
\caption{As Figure \ref{fig:zzfull} but with all objects at redshifts $1.3<z<1.7$ removed from the training set.  }\label{fig:zzdesert}
\end{figure*}

The case of removing objects in the $1.3<z<1.7$ ``redshift desert'' is a little more interesting.  Although Figure \ref{fig:deserttest} shows very little effect on RMS errors for even quite severe reductions in the number of training objects in the redshift desert, Figure \ref{fig:zzdesert} shows that, in the extreme case where all objects at these redshifts are eliminated, the behaviour of GP and ANN regression is quite different; the GP seems to much more easily interpolate across the gap in redshift than the ANN.  This is potentially a useful property when dealing with any training set with gaps in parameter space; in this case the neural network method causes a false concentration of test objects at the upper limit of the redshift desert, which would likely cause false positives in searches for clusters, for example.

Overall, we see that off-the-shelf GP regression has both advantages and disadvantages compared to {\sc{ANNz}}.  {\sc{ANNz}} makes full use of the uncertainty information for each datapoint, and thus produces reliable error estimates on the redshift, provided that the training set is reasonably large and representative.  Its ability to use uncertainty information properly may also be responsible for its higher precision in the large training-set limit; the GP regression treats all data equally, even those with large errors on their photometry.  On the other hand, the GP has far fewer hyperparameters to fit, and, perhaps because of this, appears to behave more stably in the case of sparse data.  

Since we know that GP and ANN regression have identical mathematical properties (at least for special cases of both methods, as discussed above), we hope that it should be possible to combine the best properties of both methods explored here.  This might be achieved either by incorporating a treatment of inhomogeneous errors into the GP method, or by training ANNs differently, e.g. allowing freedom in the network architecture and selecting the best network based on Bayesian evidence rather than a validation set.  Of these possibilities, the more-flexible ANN option is better developed in the machine-learning community, but the addition of input errors to GP regression ought to be feasible \citep[as discussed by][]{girard05} and may offer a simpler training option.

\section{Conclusions}
We have compared a simple implementation of GP regression, based on the {\sc{stableGP}} code of \citet{foster09}, with the popular neural network code {\sc{ANNz}}.  

With large, complete training sets the precision of the methods is very similar, with {\sc{ANNz}} achieving slightly smaller RMS errors, particularly at redshifts $z<1$. However,   we note that GP regression, unlike {\sc{ANNz}}, does not require an additional set of validation data to preserve generality, and so could take advantage of a larger training sample. 

We investigate placing magnitude limits on the training sets but find that the performance of neither algorithm is strongly affected, until the limits are so severe that they remove a large fraction of the training objects in the redshift range of the test object.  In this case the GP algorithm produces less biased results but with a larger scatter.  Similarly, the introduction of a plausible number (up to 10\%) of inaccurate redshifts into the training set, or the removal of objects of a particular spectral type, has little effect on either method.  

If the size of training set is instead reduced by random sampling, the RMS errors of both methods increase, but they do so to a lesser extent and in a much smoother manner for the case of GP regression.  With this particular dataset and network architecture we find that the ANN method deteriorates sharply with training sets smaller than about 2000 objects, where its RMS errors are $\sim20$\% larger than for GP regression, with considerably increased bias.  While the GP results could in principle be reproduced by some appropriately selected ANN architecture, the computational difficulty of selecting and then training such a network may make the GP method preferable.  

In addition, when training objects are removed at redshifts $1.3<z<1.7$, to simulate the effects of the ``redshift desert'' of optical spectroscopy, the GP regression is successful at interpolating across the redshift gap, while {\sc{ANNz}} suffers from strong bias for test objects in this redshift range.

These properties  make GP regression a very attractive algorithm for the estimation of photometric redshifts, particularly in the case where few training redshifts are available at the redshift of interest.  Since telescopes of any given aperture are in general able to image fainter objects than they can obtain spectra for, this is likely to be the situation for the vast majority of both present and future deep, high-redshift surveys.  

At present, unlike the ANN code, public GP regression codes do not take account of inhomogeneous measurement errors on the photometric data, and thus cannot estimate reliable uncertainties on the predicted redshifts.  Such uncertainty values can be used to form subsamples of objects with, on average, more accurate redshift estimates.  While such subsamples have limited applications due to the complex selection effects involved, redshift uncertainty values  can also  be used (e.g. using a Monte Carlo method) to calculate more robust statistics about the whole population of galaxies.  

The advantages of GPs demonstrated in this paper suggest that such improved GP algorithms should be pursued, and we are exploring both Monte Carlo and analytic options for improving their treatment of errors.

\section*{Acknowledgments}
We make use of simulations made available by work including Adam Amara, Peter Capak, Eduardo Cypriano, Ofer Lahav and Jason Rhodes, and are grateful to Ofer Lahav for helpful discussions.  MJJ acknowledges the support of an RCUK fellowship.  FBA acknowledges the support of a Leverhulme Trust Early Career Fellowship and a Royal Society University Research Fellowship.

\bibliographystyle{mn}
\bibliography{photoz}

\bsp

\label{lastpage}

\end{document}